\documentclass[twocolumn,aps,prc]{revtex4}

\usepackage[utf8]{inputenc}
\usepackage{amsfonts}
\usepackage{amsmath}
\usepackage{amssymb}
\usepackage{epsfig}
\usepackage{bm}

\usepackage{fixme}
\usepackage{color}
\usepackage{cancel}
\begin{document}

\title{Three-body continuum states and Efimov physics in non-integer geometry}

\author{E. Garrido$^{1}$, E.R. Christensen$^{2}$, A.S. Jensen$^{2}$}

\affiliation{$^{1}$Instituto de Estructura de la Materia, IEM-CSIC,
Serrano 123, E-28006 Madrid, Spain}

\affiliation{$^{2}$Department of Physics and Astronomy, Aarhus University, DK-8000 Aarhus C, Denmark}

\date{\today}

\begin{abstract}
Continuum structures of three short-range interacting particles in a
deformed external one-body field are investigated.  We use the
equivalent $d$-method employing non-integer dimension, $d$, in a spherical calculation with a
dimension-dependent angular momentum barrier.  We focus on dimensions
close to the critical dimension, $d=d_E$, between two and three, 
defined by zero two-body energies, where the Efimov effect can occur. 
We design for this dimension region a schematic,
long-distance realistic, square-well based, three-body spherical model,
which is used to derive analytic expressions for the wave
functions, scattering lengths, phase shifts, and elastic scattering
cross sections. The procedure and the results are universal, valid
for all short-range potentials, and for large scattering lengths.
We discuss the properties and validity of the derived expressions
by means of the simplest system of three identical bosons. 
The derived expressions are particularly useful for very small energies,
where full numerical calculations are often not feasible. For energies
where the numerical calculations can be performed, a good agreement 
with the analytic results is found. These model results may be tested by
scattering experiments for three particles in an equivalent external
deformed oscillator potential. The cross sections all vanish in the
zero-energy limit for $d<3$ with definite $d$-dependent power of
energy.
\end{abstract}

\pacs{21.45.-v, 21.60.n, ? }

%21.45.-v : Few-body systems

%21.60.n : Nuclear structure models and methods

\maketitle

\section{Introduction}                            
\label{sectI}

Different spatial dimensions are known to exhibit very different
few-body structures as exemplified by the Efimov effect
\cite{efi70,bru79,lim80,nie97,vol13,nai17}, which is present in three
but absent in two dimensions \cite{nie01}.  This fact implies that
some states must change character from bound to continuum states, or
vice versa, when the spatial dimension varies continuously between two
and three.  An analytic continuation connecting these two dimensions
is therefore interesting. This is testified by previous investigations
using methods designed to reveal the huge variation of the properties
\cite{lev14,yam15,san18,ros18,chr18}.

The continuous dimension, $d$, is directly a parameter in the
spherical $d$-method introduced in Ref.~\cite{nie01}.  The
corresponding theoretical description of non-integer dimension
connecting these limits has been developed during the last years for
two and three bound particles \cite{gar19a,gar19b,gar20,gar21}.  The
method is technically precisely as complicated as other few-body
calculations without any external field.  The simplification is achieved
by use of a centrifugal dimension-dependent barrier, which replaces,
and is equivalent to use of, a deformed external squeezing potential
in ordinary three dimension calculations.  One significant achievement
from this method was the demonstration that the Efimov effect can be
induced by an external squeezing potential acting in three dimensions
\cite{gar21}, which corresponds to a critical dimension, $d=d_E$, in the 
$d$-formalism. 

The $d$-method has so far only been applied to bound states, which, by
definition, fall off exponentially with increasing relative distance.
However, very recently the method has been extended to two-body
continuum states, where wave functions, scattering lengths, phase
shifts, and cross sections are computed \cite{gar22}.  This extension
is far from trivial, since the wave functions now extend their
oscillations to infinity in the non-confined directions.  Furthermore,
phase shifts in the spherical $d$-method involve finite, and in general
non-integer, effective
angular momentum, where the realizable deformed external field has no
angular momentum barrier.  It is then remarkable that the $d$-method
phase shift is equal to the phase difference between the two
equivalent three-dimension two-body wave functions calculated with
and without short-range interaction \cite{gar22}.

The actual interest in the present type of $d$-method investigations
is that practical up-to-date few-body calculations for cold atomic and
molecular gases are easily possible.  The measurements on such gases
involve simulation and manipulation in laboratories.  These
experimental methods are extremely flexible allowing both huge
two-body interaction variation as well as an overall confining
deformed external field
\cite{vog97,cou98,nie99,koh06,blo08,chi10,den16}.  In strong contrast
to many-body approximations \cite{hov18}, the advantage of few-body
physics is that all degrees-of-freedom are accurately treated.
Two-body theoretical physics is close to being analytic and even
universal in the large-distance limit for short-range interactions.
The universality can be used to compare properties of systems from
different subfields and therefore a tool to exchange knowledge between
science subfields
\cite{efi70,nai17,sim76,lan77,nie01,jen04,fre12,zin13,gar18}.

However, the necessary external field on two interacting particles
increases the difficulties to the non-analytic level of a three-body
problem.  This is accentuated if one particle is added to a total of
three particles, which in an external field would be equivalent to a
four-body problem.  The computational simplifications of reducing the
number of degrees-of-freedom as offered by the $d$-method is tempting.
However, the $d$-method applied to three particles is only available
for bound states, but not established for continuum properties. Thus,
so far scattering experiments cannot be analyzed.

The purpose of the present work is to repair this deficiency and report
on the extension of the $d$-method to three particles in the
continuum, that is wave functions, scattering lengths, phase shifts,
and cross sections.  Scattering, or continuum structure, is more
difficult to treat than discrete bound state properties
\cite{mes61,sha18}. Three particles may reveal new possibilities as
seen for Efimov induced bound states \cite{gar21}.  In fact, we shall
in particular focus on dimensions close to the critical value, $d_E$,
where the Efimov conditions are close to be fulfilled. In this region,
for $d\lesssim d_E$, the energy of the bound two-body subsystems is
very small, making pretty difficult to compute numerically the elastic
phase shifts of the so-called $1+2 \rightarrow 1+2$ reactions,
corresponding to an elastic collision between a particle and a bound
two-body system, for collision energies below the two-body breakup
threshold. In this case the energy window available for the projectile
is extremely small, and the correct asymptotics of the wave function
is reached at an extremely large distance, where the three-body wave
function should be computed with sufficient accuracy
\cite{gar12,gar13,bar07,rom11}.  Analytic expressions for these
phase shifts, as well as for the case of $3\rightarrow 3$ reactions,
with pure initial and final three-body states, will be provided.

These three-body systems may themselves be of practical use, but also
able to teach us how to create and control similar properties in
systems, which so far are outside experimental reach.  The paper is
organized into the introduction in Section~\ref{sectI} followed by
Section~\ref{sect2}, presenting the formalism and describing a
semi-realistic model providing analytic expressions for scattering length 
and phase shift. The properties of the three-body potentials
and the phase shifts are discussed and illustrated in Section~\ref{sect3}. 
The connection between the computed $d$-dimension wave function and
the one corresponding to the squeezed three-dimension space, which allows
to compute observables like the cross section, is discussed in Section~\ref{sect4}. 
Finally, Section \ref{sect5} contains the summary and the conclusions.

\section{Three-body continuum states}                            
\label{sect2}

 The basic method to treat three-body systems relies on the use of 
hypersperical coordinates, where several bound state investigations
are available in the literature for non-integer dimension
\cite{gar19a,gar19b,gar20}.  The hyperradius, $\rho$, is defined by
\begin{eqnarray} \label{e700}
  m M \rho^2 = \sum_{i<j}  m_i m_j (\bm{r}_{i}- \bm{r}_{j})^2  \; 
\end{eqnarray}
in terms of the radii, $\bm{r}_{i}$, and masses, $m_i$, of the three
particles, $M=m_1+m_2+m_3$, and $m$ is the arbitrary normalization mass.

Interesting structures are related to changing dimensions, as
already seen for two particles.  However, the pathological Efimov
effect for three particles, is producing numbers of bound states
varying from zero at $d=3$ over infinitely many at $d=d_E$, and back
again to a finite number for $d=2$.  The strong variation of numbers
and energies of bound states and resonances must necessarily influence
cross sections for various types of scattering between three
particles.  Investigating the corresponding $d$-dependent properties
of these states is the purpose of this section, where we use an
appropriate semi-realistic schematic large-distance model to extract
the essence. For simplicity, we shall consider a three-body system where
only relative $s$-waves enter.

\subsection{Semi-realistic model}

We seek the analytic insight, which only can be obtained by use of
schematic models, where approximations are inevitable. 
When applied to few-body systems, and in particular to three-body systems,
this kind of model, despite its simplicity, permits a simple access to their
main properties and characteristics, which furthermore are often described 
with sufficient accuracy \cite{inc05}. We shall focus
on structures arising when the Efimov conditions are approached, i.e., 
for values of $d$ in the vicinity of the critical dimension $d_E$.  In
this context, we shall first derive three-particle scattering
properties.  

The key equation is the decoupled reduced radial,
$\rho$-dependent, $s$-wave Schr\"{o}dinger equation, that is \cite{nie01,gar20}:
\begin{align} \label{e710}
  \left( - \frac{\partial^2 }{\partial \rho^2} +
  \frac{\lambda(\rho)+\ell_d(\ell_d+1)}{\rho^2} -
  \frac{2m}{\hbar^2} E \right) f(\rho) =0 \;,
\end{align}
where $\ell_d=d-3/2$ is the $d$-dependent effective angular
momentum quantum number for three particles.  We immediately emphasize
that the $\ell_d\equiv \ell_{d,N=3}$ used in Eq.(\ref{e710}) for
three particles, differs from $\ell_{d,N=2}=(d-3)/2$ for two particles.  
This use of non-integer angular momentum quantum number is similar to 
the Regge pole analytic continuation \cite{reg59}.

The hyperangular eigenvalue,
$\lambda(\rho)$, includes the short-range interaction, it is in
principle $\rho$-dependent, and it is in
general fully determined for all $\rho$, from zero to
infinity, in the $d$-dependent adiabatic hyperspherical
expansion method \cite{nie01}.  We emphasize that we consider only the
lowest adiabatic potential, assumed to be decoupled from the rest. 
This assumption is pertinent when describing low-energy phenomena,
such as the  Efimov physics. In fact, in the limit of zero energy and infinite
scattering lengths, the non-adiabatic coupling terms are identically zero.

We note that Eq.(\ref{e710}) defines a reduced radial equation
with an effective potential 
\begin{equation}
U_{\mbox{\scriptsize eff}} = \frac{2m}{\hbar^2} V_{\mbox{\scriptsize eff}}
=\frac{\lambda(\rho)+\ell_d(\ell_d+1)}{\rho^2}.
\label{efpot}
\end{equation}

The details of the short-range three-body potential are unimportant in
the cases of our interest, where the large-distance properties are
decisive.  This has two consequences.  First, the only role played
by the short-distance $\lambda(\rho)$-behavior is to deliver
the attraction required for the three-body
system.  We are free to use any schematic three-body potential for
small $\rho$, for example a square-well or its limit of a zero-range
potential. Second, for our purpose we can efficiently use the
universal large-distance structure for $\lambda(\rho)$.  Thus, we
divide the $\rho$-axis into three intervals: $0 \leq \rho \leq \rho_{0}$
(interval I), $\rho_{0} \leq \rho \leq |a_{\mbox{\scriptsize av}}|$ (interval II), 
and $\rho \geq |a_{\mbox{\scriptsize av}}|$ (interval III).
The two separating lengths, $\rho_{0}$ and $|a_{\mbox{\scriptsize av}}|$, will be defined
in the following discussion.  

The small-distance interval, $0 \leq \rho \leq \rho_{0}$, interval I, contains the
dependence on the small-distance properties of the two-body potentials
involved in the three-body system, whose main effect is to determine the energy of 
the bound three-body state, upon which the possible series of excited, may be Efimov, 
states are built. The structures of these excited states are potential-independent in
the sense that their large-distance properties would be the same for
any potential with the same scattering length \cite{gar18}.  Without
loss of generality we therefore choose a two-body square-well potential of
depth, $V_0>0$, and radius, $r_0$, which for simplicity in the notation
will be taken the same for each of the three pairwise interactions involved
in the three-body system. 

Different estimates can be made in order to fix the value of $\rho_0$ determining the 
upper limit of interval I. To do so, let us remind that the hyperradius defined in 
Eq.(\ref{e700}) can also be written in terms of the usual $\bm{x}$ and $\bm{y}$ 
Jacobi coordinates, $\rho^2=x^2+y^2$, which is independent of the Jacobi set chosen to describe the system 
\cite{nie01}, and where $\bm{x}=\sqrt{\mu_{ij}/m}\bm{r}_{ij}$ and $\bm{y}=\sqrt{\mu_{ij,k}/m}\bm{r}_{ij,k}$ 
($\mu_{ij}$ and $\mu_{ij,k}$ are the reduced masses of the $ij$ and the $(ij)k$ systems,
respectively, and $\bm{r}_{ij}$ and $\bm{r}_{ij,k}$ their corresponding relative
distances).

One possible estimate of $\rho_0$ can be made by taking it equal to the smallest of the 
three available values $\sqrt{\mu_{ij}/m}r_0$, which for three identical 
particles with mass $m$ corresponds to $\rho_0=r_0/\sqrt{2}$. This is
the value of $\rho$ obtained when $r_{ij,k}=0$ and the distance between the other two
is equal to $r_0$ (three aligned particles). This definition guarantees that for any geometry 
with $\rho < \rho_0$
the three pairs of particle-particle distances are smaller than $r_0$. However, with this
choice we disregard geometries with $\rho>\rho_0$ but still with all the interparticle
distances smaller than $r_0$. An alternative could be to proceed the other way around,
that is, to construct $\rho_0$ assuming that two of the particles are at the same position, 
$r_{ij}=0$, and take $r_{ij,k}=r_0$ (dimer-particle structure). This leads, for three identical 
particles of mass $m$,
to $\rho_0=\sqrt{2/3}r_0$, although in this case it is possible to find geometries
with $\rho\leq \rho_0$ and some particle-particle distance larger than $r_0$ 
(for instance when $\rho=\rho_0=\sqrt{2/3}r_0$ and $r_{ij,k}=0$, we get that
$r_{ij}=2r_0/\sqrt{3}$). In any case, we can conclude that an appropriate
choice for $\rho_0$ should be, for three identical particles with mass $m$, in the vicinity of 
$\rho_0=r_0/\sqrt{2}\approx 0.7 r_0$ or $\rho_0=\sqrt{2/3}r_0\approx 0.8 r_0$.

According to this, in the small-distance interval I, Eq.(\ref{e710})
reduces to the Schr\"{o}dinger-like equation \cite{nie01},
\begin{eqnarray} \label{e720}
  \bigg( - \frac{\partial^2 }{\partial \rho^2} &+&
  \frac{\ell_d(\ell_d+1)}{\rho^2} - k^2 \bigg) f_I(\rho) =0, \\ \label{e730} 
  k &=& \sqrt{\frac{2m(V_{03}+E)}{\hbar^2}} \;,
\end{eqnarray}
where $f_I$ is the related wave function in this interval, and $k$ is
the wave number related to the present three-body problem. In Eq.(\ref{e730}),
$V_{03}=3V_0$, due to the presence of 
three particle-particle interactions within the three-body system.

The crucial intermediate interval, $\rho_{0} < \rho < |a_{\mbox{\scriptsize av}}|$, interval II,
is upwards limited by $|a_{\mbox{\scriptsize av}}|$, which for three arbitrary masses can
be defined in analogy to  Eq.(\ref{e700}):
\begin{eqnarray} \label{e740}
 m M a^2_{\mbox{\scriptsize av}} = \sum_{i<j} a^2_{ij} m_i m_j  \;,
\end{eqnarray}
or perhaps as in Ref.~\cite{fed03}, 
\begin{eqnarray} \label{e749}
  a_{\mbox{\scriptsize av}} \sqrt{m} = \frac{\sqrt{2}}{3} \sum_{i<j} a_{ij} \sqrt{\mu_{ij}} \;,
\end{eqnarray}
where $\mu_{ij}$ is the reduced mass of particle pairs $i$ and $j$ and
$a_{ij}$ is the $d$-dependent scattering length of the potential between particles
$i$ and $j$.
Both definitions in Eqs.(\ref{e740}) and (\ref{e749}) have the merit
of returning $a_{\mbox{\scriptsize av}}$ as the initial two-body scattering length for
three identical particles with mass $m$.  Another definition may turn
out to be more suitable for asymmetric systems, but the limit for
identical particles must be maintained.  

The necessary assumption here is
that the supporting decoupled adiabatic potential is sufficiently
accurate for our purpose. The Schr\"{o}dinger equation is then
\begin{eqnarray} 
  \bigg( &-& \frac{\partial^2 }{\partial \rho^2} + \frac{\lambda_\infty+\ell_d(\ell_d+1)}{\rho^2}
  - \frac{2m}{\hbar^2} E \bigg) f_{II}(\rho) \nonumber \\ \label{e750}
  &=& \left( - \frac{\partial^2 }{\partial \rho^2} + \frac{-|\xi_d|^2-1/4}{\rho^2}
 - \kappa^2  \right) f_{II}(\rho)  = 0,
\end{eqnarray}
where $f_{II}$ is the relative wave function in this interval, and
the energy-related wave number, $\kappa$, is defined by
\begin{eqnarray}  \label{e753}
 \kappa^2 = \frac{2m E}{\hbar^2} \; .
\end{eqnarray}

The effective potential in Eq.(\ref{e750}) is given by the
$\rho$-independent $\lambda$-value at large-distances, $\lambda_\infty$, 
including the
angular momentum term \cite{nie01}, and before the (large) scattering
length is reached.  
The value of $\lambda_\infty$ is determined by the universal
non-analytic imaginary solution, $\xi_d$, to a transcendental equation
\cite{nie01,chr18,mik15} depending on $d$ and on the particular three-body
system under investigation.  The actual value of $|\xi_d|$ is closely
related to the well-known scaling of energies and sizes of Efimov
states \cite{efi70,gar18}.  The numerator in the centrifugal barrier
term is then $\lambda_\infty + \ell_d(\ell_d+1) \equiv \ell^*(\ell^*+1)$, 
where $\ell^* =i|\xi_d| -1/2$ is an effective complex angular momentum,
again similar to the Regge pole analytic continuation \cite{reg59}.  

This approximation of the adiabatic potential
requires a large scattering length, $|a_{\mbox{\scriptsize av}}|$, which is in complete
agreement with the potential-independence valid at distances far
outside the short-range three-body potential.  All assumptions are
very well fulfilled for dimension $d$ sufficiently close to $d_E$,
where the two-body scattering length is infinitely large.

The potential in the third interval, interval III, at large distances outside the
average scattering length, $\rho>|a_{\mbox{\scriptsize av}}|$, is again taken from the
general behavior of $\lambda(\rho)$ \cite{nie01,chr18}.  The Schr\"{o}dinger
equation is now in this interval:
\begin{eqnarray} \label{e760}
  \bigg( - \frac{\partial^2 }{\partial \rho^2} &+& \frac{\ell_C(\ell_C+1)}{\rho^2}
  - \kappa_C^2 \bigg) f_{III}(\rho)=0,  \\  \label{e770}
   \kappa_C &=& \sqrt{\frac{2m }{\hbar^2} (E - C)} \;.
\end{eqnarray}
 The
definitions of $\kappa_C$ and $\ell_C$ are related to the characteristics
of the three-body problem, and determined from the two possible
behaviors of $\lambda(\rho)$ at large distance.  The
first possibility is $\lambda(\rho) \stackrel{\rho\rightarrow \infty}{\longrightarrow} 0$ ($C=0$), which
corresponds to the case when no two-body bound state exists, and for
which we have $\kappa_C=\kappa$, Eq.(\ref{e753}), and
$\ell_C=\ell_d=d-3/2$.  The change in the effective potential
from interval II to interval III very likely implies that the 
potential is discontinuous at $\rho=|a_{\mbox{\scriptsize av}}|$, but still perfectly allowed in the equation of motion.

The second possibility is that 
$\lambda(\rho) \stackrel{\rho\rightarrow \infty}{\longrightarrow} 2 m \rho^2 E_2 /\hbar^2$,
and $C=E_2<0$, when an asymptotic two-body bound state 
structure of binding energy, $|E_2|$, is approached.  In this case
$\kappa_C\neq \kappa$, and $\ell_C=(d-3)/2$, which is the
$d$-dependent effective angular momentum quantum number associated to an ordinary
two-body collision in $d$-dimensions, namely, the collision between
the third particle and the bound dimer.

It is important to keep in mind that in this analysis we are
considering only one $\lambda$-function, which amounts to consider
only the first, usually dominant, term in the adiabatic expansion of
the wave function \cite{nie01}. This means in practice that we are restricting
ourselves to scattering processes where the incoming and outgoing
channels are described by the same $\lambda$-function. In other words,
if $\lambda \stackrel{\rho \rightarrow \infty}{\longrightarrow} 0$ we
are then dealing with a $3 \rightarrow 3$ process, where we have an unbound
three-body system in the initial and final state. On the contrary, if
$\lambda \stackrel{\rho \rightarrow \infty}{\longrightarrow} 2 m
\rho^2 E_2$, we are then dealing with an elastic scattering process,
where one of the particles hits the bound two-body state leading
to the same particle+dimer system after the collision. We shall
refer to these processes as $1+2\rightarrow 1+2$ reactions. 
For both, $3 \rightarrow 3$ and $1+2\rightarrow 1+2$ processes,
note that the dimensionless quantity $\kappa_C r_0$ is nothing
but $\sqrt{E_{\mbox{\scriptsize inc}}}$, where $E_{\mbox{\scriptsize inc}}$
is the incident collision energy in units of $\hbar^2/(2mr_0^2)$,
as one can immediately see from Eq.(\ref{e770}).

The description of other processes, like the recombination of three
particles into a particle+dimer state, or breakup 1+2 reactions,
unavoidably requires the inclusion of additional open channels.
This  necessarily requires calculation of the full ${\cal S}$-matrix
of the process \cite{gar12,gar13}.  This generalization will not be
considered here.

\subsection{Phase shifts}

The wave functions are found by solving Eqs.(\ref{e720}), (\ref{e750})
and (\ref{e760}) in their respective intervals with subsequent
matching at the boundaries, where both $\rho=0$ and $\rho=\infty$ are
included.  The distance dependent inverse square effective potentials
allow analytic solutions even though the strength sometimes is
negative in contrast to the well-known centrifugal barrier solutions.
We consider the case of continuum states, but the procedure can as
well be applied to bound state computations.  The formal expressions
for the radial solutions are:
\begin{eqnarray} \label{e780}
  f_{I}(\rho) &\propto&    \rho j_{\ell_d}(k\rho),  \\ \label{e790}
  f_{II}(\rho)  &\propto&  \cot\delta_m \rho j_{\ell^*}(\kappa \rho)
  - \rho \eta_{\ell^*}(\kappa \rho),  \\ \label{e800}
f_{III}(\rho)  &\propto&   \cot\delta_o \rho j_{\ell_C}(\kappa_C \rho)
  -  \rho \eta_{\ell_C}(\kappa_C \rho)   \;,
\end{eqnarray}
where $j_\ell$ and $\eta_\ell$ are the spherical Bessel functions with
the given indices and arguments. The indices may be continuous, like
$\ell_d$, or even complex, like $\ell^*$, and the arguments may also 
take complex values.  In any case,
the analytic continuation of the Bessel functions guarantee unique
definitions.  The precise normalization is not relevant for our
purpose. The continuous functions, $\delta_m$ and $\delta_o$, are
phase shifts as functions of energy for continuum solutions.
  We already inserted the finite
boundary condition at $\rho=0$ for $f_{I}$, since $j_{\ell_d}(k\rho)
\rightarrow 0$ for $\rho \rightarrow 0$, but $\eta_{\ell_d}(k\rho)
\rightarrow \infty$ for $\rho \rightarrow 0$.

To find the relation to the still unknown constants, the phase shifts,
we have to match the logarithmic derivatives at the two boundaries,
$\rho_{0}$ and $|a_{\mbox{\scriptsize av}}|$. The first one, the match at $\rho=\rho_0$,
leads to:
\begin{eqnarray} \label{e810}
  \frac{f_I'(\rho_{0})}{f_I(\rho_{0})} = \frac{f_{II}'(\rho_{0})}{f_{II}(\rho_{0})}  \; ,
\end{eqnarray}
where the prime means derivative with respect to $\rho$. After some algebra, 
and exploiting the well-known relations of the Bessel functions and their
derivatives, one can easily obtain:
\begin{eqnarray} \label{e820}
  \cot\delta_m =
  \frac{\kappa \rho_{0} \eta_{\ell^*+1}(\kappa \rho_{0})+ C_1 \eta_{\ell^*}(\kappa \rho_{0})}
  {\kappa \rho_{0} j_{\ell^*+1}(\kappa \rho_{0}) + C_1 j_{\ell^*}(\kappa \rho_{0})} ,  
\end{eqnarray}
where $\kappa$ is from Eq.(\ref{e753}) and the constant, $C_1$, is
given by
\begin{eqnarray} \label{e830}
  C_1 = (\ell_d-\ell^*) - k \rho_{0} \frac{j_{\ell_d+1}(k\rho_{0})}{j_{\ell_d}(k\rho_{0})}\;.
\end{eqnarray}

In the same way, the second boundary matching at  $\rho=|a_{\mbox{\scriptsize av}}|$ imposes the condition
\begin{eqnarray} \label{e850}
  \frac{f_{II}'(|a_{\mbox{\scriptsize av}}|)}{f_{II}(|a_{\mbox{\scriptsize av}}|)} =
  \frac{f_{III}'(|a_{\mbox{\scriptsize av}}|)}{f_{III}(|a_{\mbox{\scriptsize av}}|)} \; ,
\end{eqnarray}
which, similarly to Eq.(\ref{e820}), leads to
\begin{eqnarray} \label{e860}
  \cot\delta_o =
  \frac{\kappa_C |a_{\mbox{\scriptsize av}}| \eta_{\ell_C+1}(\kappa_C |a_{\mbox{\scriptsize av}}|) + C_2 \eta_{\ell_C}(\kappa_C |a_{\mbox{\scriptsize av}}|)}
      {\kappa_C |a_{\mbox{\scriptsize av}}| j_{\ell_C+1}(\kappa_C |a_{\mbox{\scriptsize av}}|) + C_2 j_{\ell_C}(\kappa_C |a_{\mbox{\scriptsize av}}|)},
\end{eqnarray}
where the constant $C_2$ is given by
\begin{eqnarray} \label{e870}
  C_2 &=& (\ell^*-\ell_C) \\ \nonumber
  &-& \kappa |a_{\mbox{\scriptsize av}}| \frac{\cot\delta_m j_{\ell^*+1}(\kappa |a_{\mbox{\scriptsize av}}|)
    - \eta_{\ell^*+1 }(\kappa |a_{\mbox{\scriptsize av}}|)}
 {\cot\delta_m j_{\ell^*}(\kappa |a_{\mbox{\scriptsize av}}|) - \eta_{\ell^*}(\kappa |a_{\mbox{\scriptsize av}}|)}\; .
\end{eqnarray}

It is important to keep in mind that the precise values of $\kappa_C$
and $\ell_C$ in Eq.(\ref{e860}), coming from the outer interval in
Eq.(\ref{e760}), depend on the existence or not of a bound
two-body dimer. When the dimer is present, we then have
$\kappa_C=\kappa_{C=E_2}\neq\kappa$ and $\ell_C=(d-3)/2\neq \ell_d$, whereas
if it is not, then $\kappa_C=\kappa_{C=0}=\kappa$ and $\ell_C=\ell_d=d-3/2$.

\subsection{Low-energy expansion}

Let us first consider the case when there is no bound dimer, and therefore
$\kappa=\kappa_C$. In this case the low-energy assumption amounts to
$\kappa \rightarrow 0$ and $\kappa_C \rightarrow 0$.

By expanding the Bessel functions around small arguments, the leading term
in the expansion for $\cot\delta_m$, Eq.(\ref{e820}), leads to:
\begin{eqnarray} \label{e840}
  \cot\delta_m \approx  - \frac{[(2\ell^*+1)!!]^2}{(\kappa\rho_{0})^{2\ell^*+1}}
  \left(\frac{1}{C_1} + \frac{1}{2\ell^*+1}\right) \; .    
\end{eqnarray}
In the expression for $C_1$, Eq.(\ref{e830}), we should insert $k \approx \sqrt{2m V_{03}/\hbar^2}$,
as obtained from Eq.(\ref{e730}) when $E\rightarrow 0$.
The expression in Eq.(\ref{e840}) can be generalized for non-integer $\ell^*$ as: 
\begin{align}
    \cot\delta_m \approx -\frac{\left[ 2^{\ell^*} \Gamma(\ell^*+1/2) (2\ell^*+1) \right]^2}{\pi (\kappa \rho_{0})^{2\ell^*+1}} \left(\frac{1}{C_1} + \frac{1}{2\ell^*+1}\right).
    \label{exp2}
\end{align}

In the same way, since $\kappa_C \rightarrow 0$, we can also perform the low-energy expansion of $\cot\delta_o$ in 
Eq.(\ref{e860}), which leads to expressions
analogous to the ones in Eqs.(\ref{e840}) and (\ref{exp2}), that is:
\begin{eqnarray} \label{e880}
 \cot\delta_o \approx - \frac{[(2\ell_C+1)!!]^{2}}{(\kappa_C |a_{\mbox{\scriptsize av}}|)^{2\ell_C+1}}
 \big(\frac{1}{C_2}+\frac{1}{2\ell_C+1} \big) \;,
\end{eqnarray}
which for non-integer $\ell_C$ takes the form
\begin{align}
     \cot\delta_o \approx -\frac{\left[ 2^{\ell_C} \Gamma(\ell_C+1/2) (2\ell_C+1) \right]^2}{\pi (\kappa_C |a_{\mbox{\scriptsize av}}|)^{2\ell_C+1}} \left(\frac{1}{C_2} + \frac{1}{2\ell_C+1}\right).
     \label{exp3}
\end{align}

Expanding again the Bessel functions in Eq.(\ref{e870}) we get for $C_2$ that:
\begin{eqnarray} \label{e890}
  C_2 &=& (\ell^*-\ell_C) \\ \nonumber
  &-&  \frac{ [(2\ell^*+1)!!]^2(2\ell^*+1)}
  {[(2\ell^*+1)!!]^2 + \cot\delta_m (2\ell^*+1) (\kappa |a_{\mbox{\scriptsize av}}|)^{(2\ell^*+1)}} \;,
\end{eqnarray}
where the expansion for $\cot\delta_m$ from
Eq.(\ref{e840}) can be inserted to give:
\begin{eqnarray} \label{e900}
  C_2 &=& (\ell^*-\ell_C) \\ \nonumber &-& \frac{(2\ell^*+1)}
 {1 - (1 + (2\ell^*+1)/C_1) (|a_{\mbox{\scriptsize av}}|/\rho_{0})^{2\ell^*+1}}  \; ,
\end{eqnarray}
which is independent of $\kappa$, and therefore on the energy as well.

In the second scenario a bound dimer exists, and the low-energy limit
then refers to a small particle-dimer collision energy, i.e., $E-E_2\rightarrow 0$.  From
 Eq.(\ref{e770}) it then immediately 
 follows that $\kappa_{C=E_2}\rightarrow 0$, which implies that
Eq.(\ref{exp3}) is still valid in this case.  Since $\kappa$ is given
by Eq.(\ref{e753}), we then have that in this low-energy limit $\kappa
\approx \sqrt{-2m|E_2|/\hbar^2}$, which is a purely imaginary number.

In the case of three identical particles with mass $m$, and a
sufficiently small value of $|E_2|$, we know that $|E_2|\approx
\hbar^2/(m a_{\mbox{\scriptsize av}}^2)$, which in the low-energy limit, leads to $\kappa
|a_{\mbox{\scriptsize av}}| \approx i \sqrt{2}$.  Therefore, in this case Eq.(\ref{e870})
becomes:
\begin{eqnarray} \label{e910}
  C_2 = (\ell^*-\ell_C)
  -i\sqrt{2}  \frac{\cot\delta_m j_{\ell^*+1}(i\sqrt{2}) - \eta_{\ell^*+1 }(i\sqrt{2})}
    {\cot\delta_m j_{\ell^*}(i\sqrt{2}) - \eta_{\ell^*}(i\sqrt{2})} \; .
\end{eqnarray}

In summary, at both thresholds we obtain $\cot\delta_o$ after
insertion into Eq.(\ref{exp3}) of the appropriate of the two
$C_2$-expressions, either Eq.(\ref{e900}) or Eq.(\ref{e910}).

The low-energy limit in Eq.(\ref{exp3}), or more precisely,
the low-energy limit of $\kappa_C^{2\ell_C+1} \cot\delta_o$,
is related to the three-body scattering length, $a_{\mbox{\scriptsize 3b}}$:
\begin{equation}
\lim_{\kappa_C\rightarrow 0} \kappa_C^{2\ell_C+1} \cot\delta_o = -\frac{1}{a_{\mbox{\scriptsize 3b}}^{2\ell_C+1}},
\label{a3}
\end{equation}
which characterizes the zero-energy scattering, and in two-body physics 
is also the measure of the cross section. It is formally related to the zero-energy limit
of the phase shift.  The small energy expansion (\ref{exp3}), implying small phase shift,
concludes that $(\kappa_C |a_{\mbox{\scriptsize av}}|)^{2\ell_C+1} \cot\delta_o \equiv
-1/A_{\mbox{\scriptsize 3b}}$, where $A_{\mbox{\scriptsize 3b}}$ is a constant independent of energy, but a
function of the dimension, $d$, the square well parameter,
$m\rho_0^2V_{03}/\hbar^2$, and the average two-body scattering length,
$a_{\mbox{\scriptsize av}}/\rho_0$, in units of $\rho_0$. From
Eq.(\ref{a3}) is then simple to see that 
$a_{\mbox{\scriptsize3b}}/ |a_{\mbox{\scriptsize av}}|   =A_{\mbox{\scriptsize3b}}^{1/(2\ell_C+1)}$.

The limiting phase shift is traditionally given as a power of the
scattering length multiplied by the wave number.  Since $\cot\delta_o$
diverges as $1/\delta_o$ in the zero energy limit, we can explicitly
define $\delta_o \rightarrow -A_{\mbox{\scriptsize 3b}} (\kappa_C |a_{\mbox{\scriptsize av}}|)^{2\ell_C+1}=-(\kappa_C a_{\mbox{\scriptsize 3b}})^{2\ell_C+1}$.
The resulting phase shift, $\delta_o$, is in both cases (existence of
bound dimer or not) approaching zero as a power of energy.  The power
is determined by $2\ell_C+1 = 2d-2$ when no dimer exists
($\kappa=\kappa_C \rightarrow 0$), and $2\ell_C+1 = d-2$ when a bound
two body state is present ($\kappa_C\rightarrow 0$ and $\kappa$ is
finite and imaginary).  

\subsection{Universality}

In Ref.~\cite{gar22} two-body systems were discussed analytically for 
non-integer dimensions and  general short-range potentials. The method was
to refer numerical calculations to use of a square-well potential, where its
radius, $r_0$, and depth, $V_0$, are found by the conditions that the
ratio between the scattering length and effective range in three dimensions is
identical to the same ratio for the given short-range potential.  In this
sense, the results provided in Ref.~\cite{gar22} for two-body systems
were found to have universal character. 
The appearance of the two-body effective range as the length unit 
to be used to get universality is consistent with additional findings
at the three-body level, where the two-body effective range is also
found to be the relevant length scale unit setting a universal value
for the three-body parameter \cite{nai14}. Furthermore, for a given potential
shape, any potential with the same value of $r_0^2 V_0$ does actually preserve 
the ratio between the scattering length and effective range.

At this point we can also wonder about the universal 
character of the expressions derived above for three-body systems.
We can perhaps expect that such a universal character, understood as  
giving rise to equal results for different two-body interactions, is also
related to the fact of producing the same ratio between the three-body scattering 
length and effective range for $d=3$, as found for two-body systems. However, this way of determining the equivalence 
between different potentials is at least not  very efficient, since it requires 
to solve for all of them the numerical three-body problem 
for $d=3$ and for very small energies, which could in itself be quite delicate.

The present investigation of the effects of an external deformed field 
turns actually the focus on non-integer dimensions close to the one, $d_E$, for 
which the Efimov effect occurs.
For this reason, even more crucial than producing the same ratio between the scattering 
length and effective range at the three-body level, 
is that the potentials, in order to be considered equivalent, should provide the same 
Borromean $d$-region
for the three-body system, or at least the same value of the critical dimension,
$d_E$, where the Efimov effect can appear. This last demand, having the same
$d_E$-value, has the advantage that it can be checked at the two-body level, since
$d_E$ is determined by the dimension at which a bound two-body state with zero 
energy is present. For this reason, in this work we shall consider that two short-range
potentials are equivalent when providing the same critical dimension, $d_E$.

This choice for the definition of equivalent two-body potentials is in some way the 
natural choice. Formally, the division of the $\rho$-space 
into three intervals, the extraction of the Schr\"{o}dinger equation applying on each of them,
and all the subsequent derivations, have been performed assuming square-well 
two-body potentials. This affects interval I, where the effective
three-body potential, Eq.(\ref{e720}), is also a square-well whose
depth, dictated by the depth of the two-body potentials, enters
explicitly in the definition of the wave number in Eq.(\ref{e730}).
However, the shape of the potential is not relevant here, since the depth,
through $k$ in Eq.(\ref{e720}), only modifies the inner part of the wave function.
What is really crucial is the size of the second interval, which 
is determined by the value of the average two-body scattering lengths,
$|a_{\mbox{\scriptsize av}}|$. This is the region where the effective
three-body potential is essentially given by the asymptotic value of the
$\lambda$-function, i.e., $\lambda_\infty$, which is potential independent, and 
which determines the scaling of the Efimov states. Therefore, no matter the shape 
of the two-body potentials, 
two different potentials having the same large scattering length for a
given dimension, will provide the same series of scaled Efimov states, 
and, therefore, they will give rise to very similar three-body wave functions. For this 
same reason, if two different two-body potentials produce the same 
critical dimension $d_E$ (infinite two-body scattering length), it is then
obvious that for any dimension close to $d_E$, both potentials will have pretty similar
large scattering lengths, and again, the corresponding three-body
wave functions will be similar as well.

The conclusion is then that, provided that we are dealing with dimensions close to $d_E$, 
the analytic expressions derived in the previous subsections can be applied
to any short-range two-body potential with critical dimension $d_E$, not just a square-well potential. 
The only thing
one has to do is to provide the correct $|a_{\mbox{\scriptsize av}}|$-value
for each dimension, and, for the case of $1+2 \rightarrow 1+2$ reactions,
also the correct binding for the bound dimer. The value of the depth, $V_0$,
and therefore the value of $V_{03}$ in Eq.(\ref{e720}),
can be estimated as an average of the depth in the inner part of the potential, 
or just the value of the two-body potential at the origin.

In any case, no matter the shape of the two-body potentials, in order to be 
able to find analytic solutions, we aim in 
interval I at an effective three-body square-well potential
in terms of the hyperradius, with depth, $V_{03}$, and a range, 
$\rho_0$. In particular, for three identical particles we have
that $V_{03} = 3V_0$. The three-body square-well radius, $\rho_0$, 
should reflect the two-body short-range properties, although at most 
this is possible to achieve on average. In this work two different 
estimates of $\rho_0$ have been proposed, which consist on 
different average radii for the geometry of three particles
in different arrangements such that the three interparticle 
distances are smaller than the range of the two-body potentials.

Once the value of $V_{03}$ and $\rho_0$ to be used in the analytic equations
is determined, the three-body properties might as well be obtained by any 
$\rho_0$, even very small and approaching zero-range, provided the corresponding 
hyperradial depth is accordingly adjusted. The relation can be taken analogous 
to the two-body result, which leads to an approximately constant value of the 
combination $\rho_0^2 V_{03}$. All these possibilities can be tested and 
corresponding dimensions can be calculated for any set of short-range two-body 
interactions.  These choices of parameters extend the applications for specific
interactions to universal properties.

\section{Results}
\label{sect3}

Although the analytic derivations in the section above are general,
it is obvious that numerical applications demand choosing specific systems. 
For the present first exploratory investigation we select the example
of three identical spinless particles. The mass of the particles is taken equal
to the normalization mass, $m$. This choice has the advantage of being the
system with fewest degrees-of-freedom, since all three two-body
interactions and mass ratios are identical.
In any case, asymmetric systems can also be studied numerically with the general
formalism developed in the previous section.

For the particle-particle interaction we choose the Gaussian potential
described in Ref.~\cite{gar21} for the case of identical particles.
As shown in Ref.~\cite{gar21}, after full numerical calculations,
this interaction is such that the three-body system is not
bound for $d=3$, but after some squeezing, for $d\approx 2.89$, the 
first bound three-body state shows up. The critical dimension $d_E$ 
for this potential is $d_E=2.75$, where a two-body bound state appears
with zero energy. This implies that for dimensions
within the range $2.75 \leq d \lesssim 2.89$ the three-body
system has Borromean character, whereas for $d < d_E=2.75$ bound two-body 
dimers are always present. For $d\approx 2.76$ the first bound three-body
excited state appears. The results to be shown later on with this potential 
will be compared with the ones obtained with the equivalent two-body square-well 
potential providing the same
critical dimension, $d_E=2.75$, as in the Gaussian case.

\subsection{Three-body potential}

Let us start by investigating the characteristics of the effective
three-body $\rho$-dependent potential entering in Eq.(\ref{e710}), 
which is given by Eq.(\ref{efpot}).
The procedure followed in order to construct this potential in the schematic
model contains two important approximations. First, only one of the adiabatic
hyperspherical potentials is used, where all the coupling terms are assumed small
and neglected. And second, the coordinate space is divided into three well-defined
$\rho$-intervals, each of them containing a very specific $d$-dependent centrifugal 
barrier.

The small-$\rho$ region, interval I, is of little importance and could as well be 
disappearingly narrow to the limit of a zero-range potential with correspondingly 
adjusted attractive strength. In this region, the effective potential
in Eq.(\ref{efpot}) becomes
\begin{equation}
U_{\mbox{\scriptsize eff}}  =  \frac{\ell_d(\ell_d+1)}{\rho^2}-\frac{2mV_{03}}{\hbar^2},
\label{prI}
\end{equation}
as extracted from Eq.(\ref{e720}).  
 
The intermediate region, interval II, is chosen to be that of the asymptotic large-distance 
behavior of the $\lambda(\rho)$ function close to occurrence of the Efimov effect,
that is, the  $\rho_0 \ll \rho \ll |a_{\mbox{\scriptsize av}}|$ region where $\lambda(\rho)=\lambda_\infty$ takes 
a constant value. Here, as seen in Eq.(\ref{e750}), the effective potential takes the form 
\begin{equation}
U_{\mbox{\scriptsize eff}}  =  \frac{-|\xi_d|^2-\frac{1}{4}}{\rho^2}.
\label{prII}
\end{equation} 

Finally, in the large distance region, interval III, the effective potential is 
given by the ordinary centrifugal barrier, but shifted by zero energy or by the energy
corresponding to the possible two-body bound-state, as determined by the value of $C$
in Eq.(\ref{e770}). This is shown in Eq.(\ref{e760}), from which we get that
in this interval we have,
\begin{equation}
U_{\mbox{\scriptsize eff}}= \frac{\ell_C(\ell_C+1)}{\rho^2}+\frac{2mC}{\hbar^2}.
\label{prIII}
\end{equation}

In order to compare the schematic effective potential described by Eqs.(\ref{prI}),
(\ref{prII}), and (\ref{prIII}), and the one obtained in a realistic three-body
calculation \cite{gar21}, we use the Gaussian potential introduced at the beginning of this
section. We choose two different dimensions, $d=2.90$ and $d=2.77$, which are,
respectively, rather far from or close to the critical dimension $d_E=2.75$.
For the Gaussian potential used, we have that $|a_{\mbox{\scriptsize av}}|=5.3$ 
and $|a_{\mbox{\scriptsize av}}|=96.2$ (both in units of the range of the two-body interaction, $r_0$)
for $d=2.90$ and $d=2.77$, respectively. This implies that the second
interval in our model, $\rho_{0}<\rho<|a_{\mbox{\scriptsize av}}|$, is pretty small
for $d=2.90$, whereas for $d=2.77$ it is sufficiently large to expect
that the $\lambda(\rho)$-function in Eq.(\ref{e710}) reaches the 
asymptotic value $\lambda_\infty=-|\xi_d|^2-(d-1)^2$ within the region, see Eq.(\ref{e750}).
In particular, we have that $|\xi_d|^2$ is equal to 0.978 and 0.879
for $d=2.90$ and $2.77$, respectively.

\begin{figure}[t]
\includegraphics[scale=0.30]{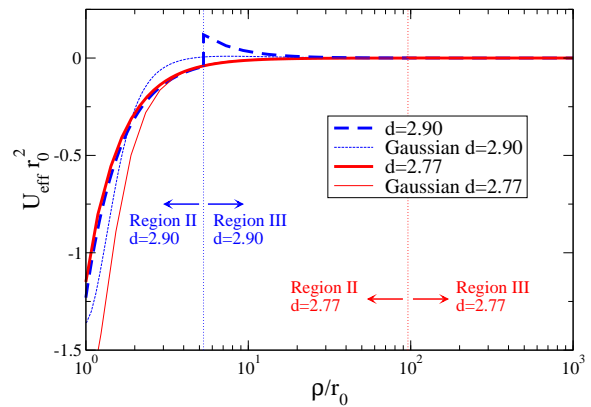}
  \caption{Three-body effective potential in intervals II and III
  for three identical bosons for $d=2.90$ (dashed curves) and $d=2.77$ 
  (solid curves). The division between intervals II and III is shown 
  by the vertical dotted lines for each of the $d$-values. The range
  to the two-body interaction, $r_0$, is taken as length unit. The mass of the 
  particles is equal to the normalization mass. The thin
  curves are full three-body calculations using the Gaussian potential 
  as described in Ref.~\cite{gar21}.
  The thick curves are the result from Eqs.(\ref{prII}) and (\ref{prIII})
  using the $|a_{\mbox{\scriptsize av}}|$-values of the same Gaussian potential.} 
    \label{3bodypot}
\end{figure}

In Fig.~\ref{3bodypot} the thin-dashed (blue) and thin-solid (red) curves
show the lowest effective potential in Eq.(\ref{efpot}), times $r_0^2$ to make
it dimensionless, obtained after a full three-body calculation, as described 
in Ref.~\cite{gar21}, for $d=2.90$ and $d=2.77$, respectively. The vertical
dotted lines indicate the value of $|a_{\mbox{\scriptsize av}}|$ for each of the two cases.
These lines determine the separation between intervals II and III in the
schematic model. As seen in Eqs.(\ref{prII}) and (\ref{prIII}), 
in these two intervals the effective potential provided by the
schematic model is dictated just by $|\xi_d|$ and $d$ (for the
dimensions chosen we have that $\ell_C=\ell_d=d-3/2$ and $C=0$),
reflecting the fact that in these two regions, II and III, the 
details of the two-body interaction are unimportant. Only in region I,
as seen in Eq.(\ref{prI}), the strength of the potential enters. However,
as already mentioned, the behavior of the effective potential in this
region has little relevance, since  our conclusions 
entirely are based on large-distance properties of the potentials.
For this reason only intervals II and III are shown in the figure. 

In the figure the thick-dashed (blue) and thick-solid (red) curves show the
result for the effective potential arising from Eqs.(\ref{prII}) and (\ref{prIII})
for $d=2.90$ and $d=2.77$, respectively. We can immediately see that,
for $d=2.90$ (dashed curves) the schematic model provides a potential
that differs clearly from the one obtained after a full calculation. 
In fact the transition between regions II and III is quite abrupt.
This is related to the fact that for this dimension the scattering length
is too small (we are far from the Efimov conditions), meaning that region II
is too small as well, and the conditions for the validity of Eq.(\ref{prII})
in such region are not properly fulfilled.

In fact, for $d=2.77$, for which the value of $|a_{\mbox{\scriptsize av}}|$ is significantly larger than the range
of the interaction, $r_0$, we can see that the computed potential, thin-solid-red, and the one 
provided by Eq.(\ref{prII}), thick-solid-red, become soon indistinguishable,
and the abrupt transition from interval II to interval III can not be
seen.

We can therefore conclude that the analytic model effective potential
and the realistic potential are almost identical for
dimensions close to $d_E$ and for hyperradii larger than $2-3$ times the
interaction radius.  For a given dimension, $d$, the potential is
essentially determined by the value of $\xi_d$.

\subsection{Phase shifts}

\begin{figure}[t]
\includegraphics[scale=0.30]{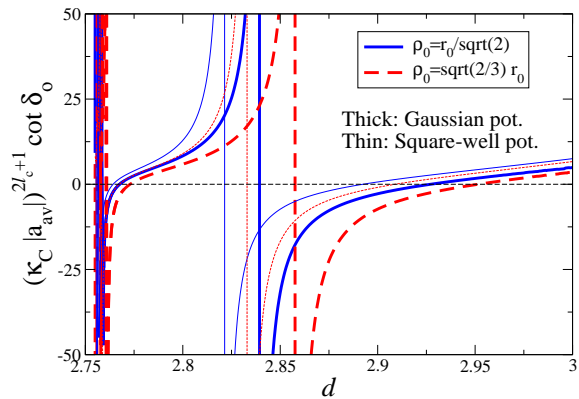}
\caption{For the Gaussian potential in Ref.~\cite{gar21} (thick curves)
and the equivalent (same $d_E$-value) square-well potential (thin curves), three-body 
value of $(\kappa_C |a_{\mbox{\scriptsize av}}|)^{2\ell_c+1} \cot\delta_o$, as function of
$d$, for $d>d_E=2.75$ and $E\rightarrow 0$, as obtained from Eq.(\ref{e860}).
The solid and dashed curves are the results obtained with $\rho_0=r_0/\sqrt{2}$
and $\rho_0=\sqrt{2/3}r_0$, respectively.}
     \label{3bcotdelt}
\end{figure}

The value of $\cot\delta_o$, which is a function of $d$, is a key 
quantity that reflects the presence of three-body bound states and
resonances, as well as for the cross sections.  More precisely,
the zeros of $(\kappa_C a_{\mbox{\scriptsize av}})^{2\ell_c+1} \cot\delta_o$ in the
$E\rightarrow 0$ limit indicate the appearance of zero energy 
bound three-body states, since these zeros correspond to infinite
three-body scattering length.

To investigate the validity of the results given in Eq.(\ref{e860}) or
(\ref{e880}), which are indistinguishable in the low-energy
limit, we show in Fig.~\ref{3bcotdelt}, as a function of $d$,
the value of $(\kappa_C |a_{\mbox{\scriptsize av}}|)^{2l+1}\cot \delta_o$ as obtained
from those equations for $E\approx 0$ and after introducing the strength, $V_{03}$,
and $a_{\mbox{\scriptsize av}}$ values corresponding to the two-body potentials.
As mentioned, we consider the Gaussian two-body potential given in
Ref.~\cite{gar21} (thick curves) and the equivalent (same $d_E$-value) square-well
potential (thin curves). In Eq.(\ref{e860}), or
(\ref{e880}), the only parameter not fully determined by the
dimension, $d$, and the interaction used, is the upper limit of interval I, i.e., the
value of $\rho_0$. Two different possibilities were
suggested in Section~\ref{sect2}, $\rho_0=r_0/\sqrt{2}$ and
$\rho_0=\sqrt{2/3}r_0$, which correspond to the solid and
dashed curves in Fig.~\ref{3bcotdelt}, respectively.

We restrict ourselves in the figure to the region $d>d_E$, since
this is the region where, when starting the confinement from $d=3$
towards smaller values of $d$, the different bound three-body states
progressively appear. In the figure we can see that when moving down 
from $d=3$, we find the first zero at $d\approx 2.93$ or $d\approx 2.95$, depending
on the value of $\rho_0$ used, for the Gaussian potential, and at
$d\approx 2.89$ or $d\approx 2.91$, for the same $\rho_0$ values, with
the square-well potential. This one should then be the dimension
for which the first three-body bound state is found. As already mentioned,
after a full three-body calculation with the Gaussian potential, as shown in Ref.~\cite{gar21}, 
we have found that the first bound state does actually appear at $d\approx 2.89$.
This value agrees reasonably well with the zero of all the functions in Fig.~\ref{3bcotdelt},
although, especially for the Gaussian case, some discrepancy is observed. 
This discrepancy is however not dramatic, since for $d\approx 2.90$, as seen 
in Fig.~\ref{3bodypot}, the schematic effective potential and the
one obtained in the full calculation clearly differ in region II,
and the schematic model is expected not to work very well. In fact,
if we keep moving down to smaller values of $d$, we observe a second
zero, and therefore the first three-body excited bound state, at a dimension
that, in all the cases, ranges between $d\approx 2.76$ and $d\approx 2.77$. 
This is as well the value quoted in 
Ref.~\cite{gar21} as the one for which the first excited state is found 
after the full three-body calculation. 

When still moving down to dimensions close to $d=d_E$, more and more, and 
eventually infinitely many, bound states appear. This leads 
to very rapid oscillations of $(\kappa_C |a_{\mbox{\scriptsize av}}|)^{2\ell_c+1}\cot \delta_o$
and to an accumulation of zeros around $d=d_E$ which are very
difficult to obtain numerically.

\begin{figure*}[t]
\psfig{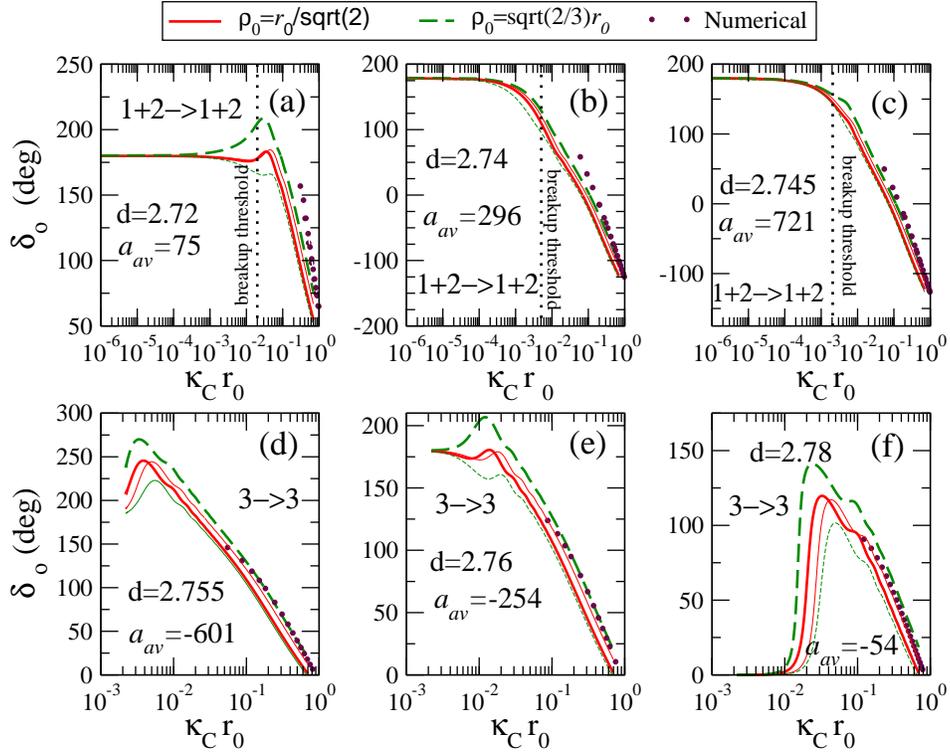}
\caption{For the Gaussian potential used in this work (thick curves) and the 
equivalent (same $d_E$-value) square-well potential (thin curves), 
three-body phase shift, $\delta_o$, in degrees, as a function of
  $\kappa_C r_0$ for different values of $d$ in the vicinity of $d_E$.  
  The solid and dashed curves are the results obtained from Eq.(\ref{e860})
  with $\rho_0=r_0/\sqrt{2}$ and $\rho_0=\sqrt{3/2}r_0$, respectively. The
  (brown) dots are the results obtained from full numerical three-body calculations
  with the Gaussian potential.  The cases in the upper and lower rows correspond to 
  $1+2\rightarrow 1+2$ ($d<d_E=2.75$) and $3\rightarrow 3$ ($d>d_E=2.75$) reactions,
  respectively. In the upper panels the vertical dotted line indicates 
  the two-body breakup threshold (very similar with both potentials). 
  The scattering length $a_{\mbox{\scriptsize av}}$ quoted in the figure for each of the cases,
  obtained with the Gaussian potential, is in units of $r_0$.}
     \label{3bdphase}
\end{figure*}

The result shown in Fig.~\ref{3bcotdelt} is very reassuring, making
us confident on the validity of the low-energy limit of Eq.(\ref{e860}),
as well as on the equivalence between the Gaussian potential and the
square-well with the same $d_E$-value in the same low-energy limit.
To investigate its validity for a larger energy range, we focus now
on the phase shifts as a function of the energy (or $\kappa_C$, Eq.(\ref{e770}))
for fixed values of the dimension. This is shown in Fig.~\ref{3bdphase},
where the different panels show the phase shift $\delta_o$ as a 
function of $\kappa_C r_0$ for several values of $d$ close to $d_E=2.75$, which
is the region where Eq.(\ref{e860}) is expected to work. 

The phase shifts are of course undetermined by a phase of 180 degrees, and
typically they are given to range either between $-90$ and 90 degrees,
or between 0 and 180. However, to avoid sudden jumps in the curves,
we have in Fig.~\ref{3bdphase} allowed the phase shifts to reach values
beyond 180 degrees or smaller than $-90$ degrees. In this way, when decreasing 
the energy, all the curves will show a smooth increase towards 180 degrees (which is equivalent 
to 0 degrees), being occasionally bigger than this value. The only exception will be panel (f), 
where the phase shifts are always within the 0 to 180 degrees range.

The upper panels
(a), (b), and (c), correspond to cases where $d<d_E$, describing
then $1+2 \rightarrow 1+2$ reactions. In these panels the vertical
dotted line indicates the $\kappa_C r_0$ value corresponding to the 
two-body breakup threshold when the Gaussian potential is used (when 
the equivalent square-well potential is used the corresponding
vertical lines can hardly be distinguished from the ones shown in 
the figure). The lower panels (d), (e) and (f),
for which $d>d_E$, describe $3\rightarrow 3$ reactions.

In the figure the thick and thin curves are, respectively, the results obtained using
the Gaussian and the equivalent square-well potentials, whereas,
for each of them, the solid and dashed curves show the results obtained from
Eq.(\ref{e860}) with $\rho_0=r_0/\sqrt{2}$ and $\rho_0=\sqrt{3/2}r_0$,
respectively. The brown dots are the results obtained from a full 
numerical three-body calculation with the Gaussian potential, where the radial wave function
associated to the lowest adiabatic potential is calculated for the
different $\kappa_C r_0$ values up to a distance of $\rho_{\mbox{\scriptsize max}} \approx 500 r_0$. 
After fitting this wave function with the asymptotic behavior given in 
Eq.(\ref{e800}) the value of $\delta_o$ is then extracted. For
$\kappa_C r_0$ values in the vicinity or smaller than about 0.1 the
correct asymptotic behavior of the numerical wave function is reached beyond
the chosen value of $\rho_{\mbox{\scriptsize max}} \approx 500 r_0$,
and the extraction of the phase shift would require a more accurate 
three-body calculation. In any case, especially for $1+2 \rightarrow 1+2$ reactions
below the two-body breakup threshold this numerical procedure is very
inefficient and very likely even numerically impossible, being
then necessary to implement an alternative \cite{bar07,rom11}.

As we can see in the figure, for both, $1+2 \rightarrow 1+2$
(upper panels) and $3 \rightarrow 3$ (lower panels) reactions, 
the agreement between the numerical phase shifts (brown dots),
in the energy region where they can be easily computed, and the 
analytic calculations from Eq.(\ref{e860}) is remarkably good.
This is especially true when choosing $\rho_0=\sqrt{2/3}r_0$ 
(dashed curves), which then appears as a better choice for the
upper limit of interval I. The other choice, $\rho_0=r_0/\sqrt{2}$
(solid curves),
is perhaps too restrictive, eliminating from interval I a bit
too much of the three-body geometries containing the three
particles within the potential range. This result is therefore 
consistent with the spatial structure of Efimov states, that is a 
coherent superposition of the three dimer-particle configurations, 
as also found in detailed analyses investigating their spatial 
structure \cite{efi73,nai14a,gar21b}.

Also, the results obtained from the Gaussian potential (thick curves) 
and the equivalent square-well potential (thin curves) are very
similar, often indistinguishable from each other. This fact supports
the universal validity of Eq.(\ref{e860}) after defining the equivalent potentials
as those giving rise to the same critical dimension, $d_E$. This
definition actually makes particular sense if we remind that Eq.(\ref{e860})
has been designed to work for dimensions in the vicinity of $d_E$.
In fact, the largest discrepancies between the different curves in 
Fig.~\ref{3bdphase} are observed in panels (a) and (f), which are as well
the cases for which the dimension, $d=2.72$ and $d=2.78$, differs
the most from $d_E=2.75$. 

The conclusion is then that the semianalytic procedure described
in this work provides reliable results for the phase shifts
in the region close to $d=d_E$, where the Efimov conditions are
close to be fulfilled. This is like this for both, $1+2 \rightarrow 1+2$
and $3 \rightarrow 3$ reactions, and the phase shifts can be 
obtained even for extremely small energies. This is particularly
relevant for elastic $1+2 \rightarrow 1+2$ collisions below the two-body
breakup threshold, since, due to the small dimer energies for $d\lesssim d_E$,
the allowed collision energies, or the $\kappa_C r_0$ values,
are extremely small as well, which makes the full numerical calculation
very delicate.

\section{Application in three dimensions}
\label{sect4}

The $d$-method has been presented as an efficient procedure to
describe a system subject to the presence of a confining external
potential.  The great advantage is that the deformed external field
is replaced by a dimension dependent angular momentum barrier in
spherical calculations.  However, in order to compare with the
available experimental data, for instance with measured cross
sections, it is necessary to translate the $d$-dimension wave
functions into the ordinary three-dimension, squeezed, space. This
is done by directly interpreting the $d$-wave function as a wave
function in three dimensions, but deformed along the squeezing
direction \cite{gar19a,gar19b,gar20}.

\subsection{Scaling and external field}

The starting point in all the derivations in Section~\ref{sect2} 
is Eq.(\ref{e710}), which amounts to assuming that only one, decoupled,
adiabatic channel is enough to describe the system. For large values of the
hyperradius, $\rho$, this equation becomes Eq.(\ref{e760}), which 
formally describes a one-body problem with
angular momentum, $\ell_C$, where initial and final states correspond to
infinitely large $\rho$.  The underlying structure of three particles
does not enter in the calculations, but only through the subsequent
interpretation, where the relative coordinate, $\rho$, is the
connection.

As mentioned, the inclusion of just one channel implies that only elastic processes
can be described within the model, that is, the incoming and outgoing
channel are the same. In other words, only two processes are possible.
The first one corresponds to elastic scattering of
one of the particles on the other two in a bound state, that is $1+2 \rightarrow 1+2$ reactions.  
The second one is the theoretical construction of simultaneous elastic scattering of
three particles in a continuum state described by one decoupled
adiabatic hyperspherical potential, that is $3 \rightarrow 3$ reactions. 
These cases correspond to the large-distance Schr\"{o}dinger equation in
Eq.(\ref{e760}), where $\kappa_C$ is from Eq.(\ref{e770}), and 
$(\ell_C,C)$ is given by $(\ell_C=(d-3)/2,C=E_2)$ or $(\ell_C=\ell_d=d-3/2,C=0)$ for 
the first and the second case, respectively. The interactions are contained in the
Schr\"{o}dinger equations for the intermediate and short-distance intervals.

Following what is done in Ref.~\cite{gar22} for two-body systems, and taking the
$z$-axis along the squeezing direction, we interpret the hyperradius $\rho$ in the
$d$-space, as the hyperradius in the three-dimension space, $\tilde{\rho}$,
but deformed along the $z$-axis by means of a scale parameter, $s$ (see \cite{gar20}
for details):
\begin{equation}
\rho \rightarrow \tilde{\rho}=\sqrt{\rho_x^2+\rho_y^2+\frac{\rho_z^2}{s^2}}=
\sqrt{\rho_\perp^2+\frac{\rho_z^2}{s^2}}.
\end{equation}

The connection between the dimension, $d$, and the scale parameter, $s$, was in Ref.~\cite{gar20}
estimated to be given by:
\begin{equation}
\frac{1}{s^2}=\left[1+\left( \frac{(3-d)(d-1)}{2(d-2)} \right)^2 \right]^{1/2},
\end{equation}
which is based on the assumption of harmonic oscillator particle-particle interaction \cite{gar20}.

In this way the computed $d$-dimension wave function, $\Psi_d(\rho)$, can be 
understood as a wave function in the three-dimension space, $\tilde{\Psi}(\rho_\perp,\rho_z,s)$.   As done in \cite{gar20}, $\tilde{\Psi}$ can then be expanded in terms of some convenient three-dimension orthonormalized basis set, 
which, in general, could depend on the hyperradius, $\rho=\sqrt{\rho_\perp^2+\rho_z^2}$, and the usual five hyperangles, $\Omega$:
\begin{equation}
\tilde{\Psi}(\rho_\perp,\rho_z,s)=\frac{1}{\rho^{5/2}}
\sum_n \tilde{f}_n(\rho,s) \Phi_n(\rho,\Omega).
\label{expan}
\end{equation}

From the expression above it is simple to extract the radial wave functions
in the three-dimension (squeezed) space, which are given by:
\begin{equation}
\tilde{f}_n(\rho,s)=\rho^{5/2} \int d\Omega \tilde{\Psi}(\rho_\perp,\rho_z,s) 
\Phi_n^*(\rho,\Omega),
\label{defrad}
\end{equation}
where $d\Omega$ is the usual phase space associated with the hyperangles for 
three particles in three dimensions and $n$ numbers the different terms
of the basis set, $\{ \Phi_n \}$. This same procedure was used in 
Ref.~\cite{gar20} for the case of bound states.

Calculation of the radial wave functions (\ref{defrad}) is particularly simple
when, as done in this work, only $s$-waves enter in the 
$d$-dimension wave function, and only the lowest adiabatic channel
is considered. If, furthermore, we choose the basis set, $\{ \Phi_n \}$,
as the one formed by the usual hyperspherical harmonics, the two integrals
in Eq.(\ref{defrad}) involving the azimuthal angles of the Jacobi coordinates
can be done analytically, and only the three remaining hyperspherical angles
have to be integrated away numerically. Even more, for large values of $\rho$,
the radial part contained in $\tilde{\Psi}$ can be replaced by the known
asymptotic form in Eq.(\ref{e800}), although, as discussed in \cite{gar22},
in order to guarantee the confinement along the $z$-axis, it 
 has to be multiplied by a factor $e^{-\rho_z^2/(2b_{ho}^2)}$. Here,
$b_{ho}$ is the harmonic oscillator length associated to the squeezing potential, 
which can be related to the dimension $d$ as given in \cite{gar20,gar22}.

Also, each term $n$ of the hyperspherical
basis set has associated a value $K$ of the hypermomentum, in such a way that the
asymptotic behavior of the radial wave functions in Eq.(\ref{defrad}) should,
in principle, take the form:
\begin{equation}
\tilde{f}_n(\rho,s) \stackrel{\rho\rightarrow \infty}{\longrightarrow}
\cot\delta_n \rho j_{K+\frac{3}{2}}(\kappa_C\rho)-\rho \eta_{K+\frac{3}{2}}(\kappa_C\rho),
\label{asy}
\end{equation}
from which the three-dimension phase shift $\delta_n$ can be extracted.

As discussed in \cite{gar22}, the asymptotic behavior in Eq.(\ref{asy})
corresponds to the usual, non-deformed, three-dimension space. Therefore the phase shift, 
$\delta_n$, obtained from it, should not necessarily
be the one in the squeezed space. In fact, without 
any interaction between the particles, the phase shift obtained
in the $d$-formalism is trivially equal to zero ($\delta_o=0$)
and the corresponding $d$-dimension radial wave function is just
$\rho j_{\ell_C}(\kappa_C \rho)$, which means $\tilde{\Psi} 
\propto \tilde{\rho}j_{\ell_C}(\kappa_C \tilde{\rho})$. Comparison of the
radial wave functions then obtained through Eq.(\ref{defrad}) and the asymptotic
behavior (\ref{asy}) gives rise to a non-zero phase shift, $\delta_{\mbox{\scriptsize free}}$.
Therefore, the correct phase shift in the confined space, $\delta_{\mbox{\scriptsize conf}}$,
should be the difference between the computed phase shift, $\delta_n$, and the
one corresponding to the free case, $\delta_{\mbox{\scriptsize free}}$. In
other words, we have that:
\begin{equation}
\delta_{\mbox{\scriptsize conf}}=\delta_n-\delta_{\mbox{\scriptsize free}}.
\label{dconf}
\end{equation}

\subsection{The $1+2 \rightarrow 1+2$ reactions}

In \cite{gar22} we have shown that,
for two-body reactions, the phase shifts obtained with the $d$-method
are the same as the phase shift, $\delta_{\mbox{\scriptsize conf}}$, obtained
as described above. The consequence is that the phase shifts obtained with the $d$-method
can be directly used in the usual expressions for elastic cross sections.
This is particularly relevant for 
$1+2 \rightarrow 1+2$ reactions, since they are formally identical to 
a two-body problem involving simply the projectile and the bound dimer.
As a consequence, we can express the low-energy elastic $s$-wave cross section, $\sigma$, 
for this kind of processes as:
\begin{eqnarray} \label{e930}
 \sigma_{1+2\rightarrow 1+2} = 
  \frac{4\pi}{\kappa_C^2} \frac{m}{\mu} \sin^2\delta_o \;,
\end{eqnarray}
where the factor $m/\mu$ enters due to the fact that $\kappa_C$ in Eq.(\ref{e770})
is defined in terms of the normalization mass $m$, whereas in two-body
reactions the correct definition of the momentum should be  
in terms of the reduced mass, $\mu$, of the projectile-target system.
Note that, in this way, Eq.(\ref{e930}) is, as it should, independent
of the arbitrary choice of the normalization mass.

For $1+2 \rightarrow 1+2$ reactions and collision energies below 
the breakup threshold (or the lowest two-body excitation energy), it is 
strictly correct that only the lowest adiabatic channel is open.  Inclusion of just this
open channel provides in general a good approximation, although one or more 
closed channels may be necessary to get very precise results \cite{bar07,rom11}. 
In such cases the low-energy cross section is well approximated by 
Eq.(\ref{e930}) \cite{gar18}. For energies above the breakup threshold,
additional (breakup) channels, in principle infinitely many, are also
open, and the phase shift corresponding to the elastic channel could
be significantly affected by the now open breakup channels. However,  
for energies still sufficiently close to the breakup
threshold, although above, the inelasticity parameter of the collision (which measures
the weight of the elastic channel) can still be pretty close to 1, and the
cross section in Eq.(\ref{e930}) can still be a good approximation \cite{gar12}.

In the low energy limit, and close to the Efimov region (large
scattering lengths), the phase shift $\delta_o$ is given in 
Eq.(\ref{exp3}) after insertion of the appropriate of the two
$C_2$-expressions, which for $1+2 \rightarrow 1+2$ reactions
is Eq.(\ref{e910}).  Since $\sin^2 \delta_o \approx \delta^2_o \propto (\kappa_C
|a_{\mbox{\scriptsize av}}|)^{4\ell_C+2}$, the threshold behavior of the cross section is
$\sigma \propto a_{\mbox{\scriptsize av}}^2 (\kappa_C |a_{\mbox{\scriptsize av}}|)^{4\ell_C} $. The two-body
scattering length then appears to be the unit scale and the approach
to zero energy corresponds to scattering for a finite angular
momentum, $\ell_C$.

\begin{figure}[t]
\psfig{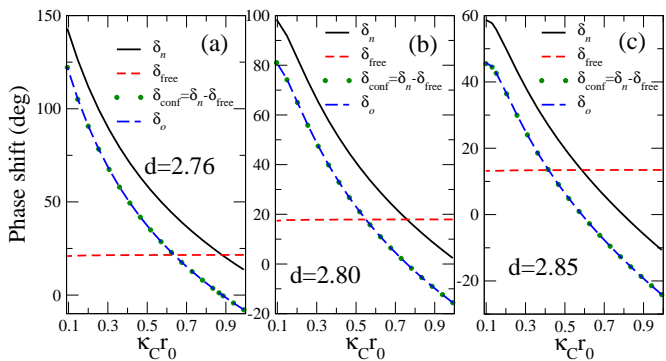}
\caption{For the Gaussian potential used in this work, and three different
values of $d$ ($d>d_E$), we give, as function of $\kappa_Cr_0$,
the phase shifts $\delta_n$ (solid-black), $\delta_{\mbox{\scriptsize free}}$ (dashed-red), 
$\delta_{\mbox{\scriptsize conf}}$ (dot-green), and $\delta_o$ (long-dashed-blue),
obtained after full three-body calculations.}
     \label{deltd}
\end{figure}

\subsection{The $3\rightarrow 3$ reactions}

The computed phase shifts, $\delta_{\mbox{\scriptsize conf}}$, in the
confined three-dimension space are given by Eq.(\ref{dconf}), which
for two-body processes have been found to be the same
as the phase shifts obtained in the $d$-formalism. It is 
interesting to note that this equality, $\delta_{\mbox{\scriptsize conf}}=\delta_o$,
still holds for pure three-body processes, like $3\rightarrow 3$ reactions.

As an illustration we show in Fig.~\ref{deltd}, as a function of $\kappa_Cr_0$, 
the values of $\delta_n$ (black solid), $\delta_{\mbox{\scriptsize free}}$ (red dashed),
and $\delta_{\mbox{\scriptsize conf}}$ (green dots), for three different dimensions in 
the $d>d_E$ region,
$d=2.76$ (a), $d=2.80$ (b), and $d=2.85$ (c). These phase shifts have been 
obtained fully numerically from a three-body calculation with the Gaussian
potential, as described for the brown dots when discussing Fig.~\ref{3bdphase}.
The phase shifts $\delta_n$ and $\delta_{\mbox{\scriptsize free}}$ are obtained
from Eqs.(\ref{defrad}) and (\ref{asy}) with and without interaction
between the particles, respectively. As mentioned, for these numerical
calculations it is not simple to reach $\kappa_C r_0$ values much smaller
than about 0.1.

In all the three cases shown in the figure, the
trend of the phase shifts is similar, and also, in all the three cases the curve
corresponding to $\delta_{\mbox{\scriptsize conf}}$ (green dots) perfectly
overlaps with the one showing the phase shifts, $\delta_o$, obtained in
the $d$-formalism (blue long-dashed). The results shown in the figure have
been obtained using the first term, $n=1$, in the expansion in Eq.(\ref{expan}),
which corresponds to $K=0$ in Eq.(\ref{asy}). We have checked that the value of 
$\delta_{\mbox{\scriptsize conf}}$ is independent of $n$, and of course, the
result $\delta_{\mbox{\scriptsize conf}}=\delta_o$ holds no matter the
$n$-term used for the calculation. 

Therefore, for $3\rightarrow 3$ reactions, the phase shifts obtained with 
the $d$-method can also be directly used in the expressions for the cross
sections. In particular, considering only the lowest adiabatic elastic
channel, the $s$-wave cross section takes the form \cite{gar18}:
\begin{equation}
\sigma_{3\rightarrow 3}=\frac{128\pi^2}{\kappa_C^5} \sin^2\delta_o,
\label{3to3}
\end{equation}
where, again, for low energies and large two-body scattering 
lengths, Eq.(\ref{exp3}) can be employed.

The cross section above presents the deficiency of actually depending
on the choice made for the normalization mass, $m$, which is contained
in $\kappa_C$, as shown in Eq.(\ref{e770}). This is related to the
fact that the incoming flux of particles is not well defined when two incident
momenta are involved. In order to avoid this problem, it is common for this
kind of reactions to deal, not with cross sections, but with reaction
rates, which are well defined. In particular, as shown in \cite{gar18}, 
the $s$-wave reaction rate for $3\rightarrow 3$ processes is given by:
\begin{equation}
R_{3 \rightarrow 3}=\frac{32\pi^2}{E^2} \hbar^5 
\left(\frac{m_1+m_2+m_3}{m_1 m_2 m_3} \right)^{\frac{3}{2}} \sin^2\delta_o.
\label{rr33}
\end{equation}

It is important to keep in mind that for $3\rightarrow 3$ processes,
and even for very low collision 
energies close to zero, it is not very realistic to assume only one open channel.  
The other channels related to the other adiabatic three-body potentials may also
contribute substantially to these cross sections \cite{gar18}.  This
is accentuated close to the Efimov threshold, where a number of bound
three-body states are present with extremely small binding energies.
All these channels are then open allowing coupling to related
inelastic scattering. To distinguish these processes must be extremely
difficult in experiments. In the present theoretical formulation this
scattering problem would also be tremendous, but it could perhaps inspire to
perform proper average accessible to experimental tests. Therefore,
the expression in Eq.(\ref{3to3}), or in Eq.(\ref{rr33}), should be taken as a very 
first approximation to the correct cross section, or reaction rate.

\section{Summary and conclusions}
\label{sect5}

In this report, we use the $d$-method to study three short-range
interacting particles in the continuum in a deformed external potential. The
$d$-formulation using one decoupled potential in the hyperspherical
adiabatic expansion leads to a radial Schr\"{o}dinger-like equation 
formally identical to the one of two particles. The hyperradius is the 
crucial distance-coordinate
describing the average distance between three particles.  They may
start being infinitely far apart, then moving together and, after
interacting, again moving away from each other.  This is either the $3$
to $3$ scattering or an elastic scattering process of $1+2$ to $1+2$,
where $2$ means that two of the three particles are in a bound state.
The formalism and technique are the same for two particles as well as
for both these cases, but the potentials differ from each other.

We insist on applications including the particularly interesting
phenomenon known as the Efimov effect.  With this in mind, we
construct an analytically solvable three-body model, where the key
quantity is the effective three-body potential.  The details of the 
short-distance properties are unimportant, and we could as well use a 
zero-range potential. More precisely, we use a square-well in 
$\rho$-space with a very short radius leaving the square-well parameter 
to be adjusted to produce the Borromean region of a more realistic potential.

The intermediate region extends from the square-well radius to the two-body
scattering length for identical bosons, or to a specified average of two-body 
scattering lengths for non-equal particles. When the scattering
lengths are sufficiently large, the shape and strength of the effective
three-body potential in this region is dictated by the large-distance (but
still smaller than the scattering length) asymptotic centrifugal barrier 
structure, which can be obtained from a transcendental equation
without any knowledge of the potentials. This amounts to use of a
generalized complex angular momentum.  This potential can also be
found from Gaussian two-body potentials, and then of a different, but
similar, structure which require numerical solution.

The last coordinate interval is beyond the two-body scattering length.
The $d$-dependent angular momentum barrier remains for $d$-values
larger than $d_E$, where the two-body subsystems are unbound.  For
$d$ smaller than $d_E$, the large-distance structure of two bound
particles is possible and the adiabatic potential can be translated by
the corresponding binding energy, still maintaining the formulation.

We calculate analytically the continuum states in this model as well
as the related scattering phase shifts, from which the low-energy
threshold behavior and cross sections are derived.  The generalized
$d$-dependent angular momentum quantum number provides the energy
power of the phase shift approach as the energy converge towards zero.
We have compared the results obtained from the analytic expressions using
the parameters of a Gaussian and a square-well potential that give rise to
the same critical dimension, $d_E$, which provides a pretty similar Borromean 
region for the three-body system in both cases. The results are to a large extent
potential independent when the dimension is close to $d_E$.
Numerical results obtained directly from two-body potentials are virtually
impossible to obtain in this dimension region for very small energies, 
as demanded for instance in $1+2 \rightarrow 1+2$ elastic reactions,
since the energies of the bound two-body states in the 
region close to the Efimov point are, by definition, extremely small.
The analytic results then become the only realistic way of studying
these reactions. For sufficiently large energies, where numerical calculations
are more accessible, the agreement with the analytic results is remarkable. 

The $d$-formulation with spherical potentials is substantially
simpler by using a conserved generalized angular momentum quantum
number.  The external deformed field necessarily involves large numbers
of partial waves.  However, to compute observable cross sections a
translation to the complicated external field formulation in three
dimensions is necessary.  We have shown numerically that the $d$-method 
phase shift is identical to the difference between external field phase shifts with
and without short-range interaction.  In other words, the $d$-method phase shifts are
the ones in the confined three-dimension space, and they are
then directly describing the scattering between three particles.  In
particular the low-energy cross section can be obtained with the
analytic expressions for the $d$-phase shift derived in this work, and
they vanish with a $d$-dependent power of energy.

In conclusion, we have investigated and illustrated three-body scattering
processes in a $d$-dimension space, equivalent to a confined three-dimension
space, by use of an analytic schematic model.  The cross section behavior
and the insight obtained are universal, that is independent of details of
the employed short-range potentials.  The translation from $d$ to
external field is necessary, available and at least a semi-accurate
description.  The perspective in our investigations is that scattering
between particles confined by deformed external fields may be useful
tools in investigations of for example structures related to Efimov
physics. Transitions between other dimensions may also be of interest.

\begin{acknowledgements}
This work has been partially supported by the Ministerio
de Ciencia e Innovaci\'{o}n MCI/AEI/FEDER,UE (Spain) 
under Contract No. PGC2018-093636-B-I00.
\end{acknowledgements}

\end{document}